\documentclass[prd,aps,twocolumn,showpacs,preprintnumbers,amsmath,nofootinbib,amssymb]{revtex4}
\usepackage[dvips]{graphicx} 
\usepackage{amsmath} 
\usepackage{amssymb} 
\usepackage{verbatim} 
\usepackage{graphicx}% Include figure files 
\usepackage{dcolumn}% Align table columns on decimal point 
\usepackage{bm}% bold math 
\def\be{\begin{equation}} 
\def\ee{\end{equation}} 
\def\bea{\begin{eqnarray}} 
\def\eea{\end{eqnarray}} 
 
\begin{document} 
 
%\preprint{} 
 
\date{\today} 
 
\title{Angular Power Spectrum of B-mode Polarization from Cosmic String Wakes} 
 
\author{Robert Brandenberger, Nick Park and Grant Salton 
\email[email: ]{parknic, gsalton, rhb@physics.mcgill.ca}}
 
\affiliation{Department of Physics, McGill University, 
Montr\'eal, QC, H3A 2T8, Canada}

 
\begin{abstract} 

Cosmic string wakes produce direct B-mode polarization at leading
order in cosmological perturbation theory, as worked out in a previous 
publication \cite{BDH} in the case of a single string wake. Here we
compute the angular power spectrum of B-mode polarization from a
scaling distribution of string wakes. We find that the averaging
which enters in computing the power spectrum renders the signal,
which is distinctive in position space maps, too small in amplitude
to be detectable with the first generation B-mode surveys. In addition,
the spectral shape is similar to that of gravitational lensing, making
it additionally difficult to detect the cosmic string signal from the
angular power spectrum. Hence, a more promising way to constrain
cosmic strings using B-mode polarization is by analyzing position
space maps using novel algorithms such as the Canny algorithm. 

\end{abstract} 
 
\maketitle

\newcommand{\eq}[2]{\begin{equation}\label{#1}{#2}\end{equation}} 
 
\section{Introduction} 

Cosmic microwave background (CMB) polarization is emerging as new 
observational window to probe the primordial universe. CMB polarization
is produced by Thompson scattering of the CMB off of free electrons.
Anisotropies of CMB polarization are produced by the scattering of
the CMB quadrupole off of density inhomogeneities which contain
free electrons (see \cite{White} for an introduction to the physics of
CMB polarization). Polarization is described by the
polarization tensor which can be decomposed into two modes - E-mode
and B-mode. E-mode polarization has already been observed
(see e.g. \cite{Emodeobs}), and several dedicated experiments to
probe polarization at improved sensitivity are in construction
or in planning (see e.g. \cite{Future}) which will reach the required 
sensitivity to detect primordial B-mode polarization. In fact,
very recently the B-mode polarization signal induced by
gravitational lensing has been observed for the first time \cite{Hanson}.

Of specific interest here is B-mode polarization. In
cosmologies with Gaussian adiabatic fluctuations, no B-mode polarization
arises at linear order in cosmological perturbation theory. B-mode
polarization on large angular scales can be generated by primordial
gravitational waves, and on smaller angular scale by lensing of 
E-mode polarization \cite{lensing}. The search for B-mode polarization
has in particular attracted a lot of attention because of the promise
to use the results to detect primordial gravitational waves such as those
produced by a period of inflation in the very early universe \cite{Starob}
(but see \cite{holy} for a discussion of other sources of gravitational
waves in the early universe).

In this paper we study the angular power spectrum of CMB polarization
(in particular B-mode polarization) induced by a scaling solution of
cosmic string wakes. Cosmic strings (see \cite{ShellVil, HK, RHBrev}
for reviews on cosmic strings) are linear topological defects which
are predicted to form during a phase transition taking place in the very
early universe. They arise in a wide set of quantum field theory models
beyond the ``Standard Model" of particle physics. The important
point is that if our microphysics is described by any model which contains
cosmic strings, a network of such strings will inevitably form \cite{Kibble} 
in a symmetry breaking phase transition and persist to the present time. 
Cosmic strings carry energy and hence lead to gravitational effects on space-time 
which can be searched for in cosmological observations. Finding evidence for
cosmic strings would be an exciting discovery, but even non-observation
of effects predicted by strings would be interesting since it would
provide new constraints on particle physics beyond the ``Standard Model".
In fact, searching for strings in cosmological observations is a way
to probe particle physics which is complementary to accelerator tests
in the sense that the cosmological tests will probe particle physics
at the high energy end of the current region of ignorance whereas
accelerators provide probes at the low energy end.

Cosmic strings are characterized by their mass per unit length $\mu$
which is usually given in terms of the dimensionless quantity $G \mu$
(where $G$ is Newton's gravitational constant, and we use natural
units in which the speed of light $c$ and Planck's constant are set
to one). At the present time, the best constraint on $G \mu$ comes
from analyses of the angular power spectrum of CMB temperature
maps and yields the constraint $G \mu < 10^{-7}$ \cite{Dvorkin, PlanckCS}
(see also \cite{previous} for previous limits based on older
data) \footnote{Note that the exact value of the upper bound depends on 
features of the distribution of the string network which are drawn from 
numerical string evolution simulations and have systematic uncertainties 
attached to them.}. Other windows to probe cosmic strings include
high redshift galaxy surveys (see \cite{Francis} for a recent study
and for references to previous works) and 21cm redshift surveys
\cite{Oscar} (see also \cite{previous21cm}) \footnote{See
\cite{RHBCSrev} for a review on how to use new observational windows to
probe for the possible existence of cosmic strings.}. In this paper, we will
focus on signatures of strings in CMB polarization maps. In
contrast to earlier work \cite{CMBpolprevious}, we here focus
on the effects of strings between the surface of last scattering
and the present time.

Cosmic strings are either infinite or else loops. Causality tells us that
a network of infinite strings must be present at all times after the
phase transition which generates them \cite{Kibble}. We can view the network
of infinite string as a system of random walks with mean step length
(i.e. curvature radius) and separation given by a correlation length
$\xi(t)$ \footnote{We are making use of the ``one-scale" model of
the string network where the curvature radius and mean separation
of the long strings are the same \cite{onescale}.}.
Causality tells us that the correlation length must be smaller or
equal to the horizon distance $t$ \cite{Kibble}, and a Boltzmann equation which
describes the energy loss of the long string network due to
expansion and string loop production indicates that $\xi(t)$
cannot be much smaller than $t$ (see e.g. \cite{RHBrev}). 
In this paper we will focus on the signatures of the long strings.

Due to their tension, cosmic strings acquire relativistic velocities. 
Due to the conical geometry of space perpendicular to a
straight string segment (see the review in the following section),
such a string segment moving with velocity $v_s$ will produce a
wedge behind it within which the initial density is twice the background 
value. This is the string wake \cite{wake}. 

Since they are overdense regions of matter (including a fraction
of free electrons), string wakes are sites of enhanced Thompson
scattering and hence lead to CMB polarization anisotropies.
The special geometry of a string wake in position space leads
to a distinctive signal in CMB polarization maps \cite{BDH}
\footnote{See also \cite{Seljak} for earlier work on CMB polarization
from cosmic defects.}.
In particular, a string wake will produce direct B-mode polarization
with an amplitude which is statistically (averaged over all wake
orientations relative to the CMB quadrupole) equal to that of the
induced E-mode polarization. Hence, since at linear order in
cosmological perturbation theory Gaussian adiabatic fluctuations
do not induce B-mode polarization, searching for B-mode CMB
polarization appears as a promising window to detect or constrain 
cosmic strings.

As in much of the analytical work on cosmic strings and
structure formation (see e.g. \cite{RHBCSrev} for a recent
short review), we will be working in terms of a toy model \cite{onescale}
in which we break up the long string network into a set of
straight segments of length $\xi(t)$ (the typical curvature
radius of the long strings) which each live for
one Hubble expansion time (the typical time interval between
intersections of long string segments). The number density $n$ of string
segments per Hubble volume at time $t$ is fixed by the scaling 
solution
\be
n \, = \, n_w t^{-3} \, ,
\ee
where $n_w$ is a constant which according to numerical 
simulations of cosmic string networks is in the range
$1 < n_w < 10$ \cite{CSsimuls}. The set of string segments
are taken to be uncorrelated in different Hubble time steps.

In a previous paper \cite{BDH}, the position space signal of
a single cosmic string wake was studied. In this paper we
work out the angular power spectrum of a scaling distribution
of string wakes. This means, in particular, that we must
sum the contributions over all Hubble time steps $t_i$ when
string wakes are formed, and over all times $t$ when our
past light cone intersects the string wake. 

Since the string signals are characterized by
edges in position space maps, good angular resolution is
more important than full sky coverage. Hence, having
in mind application to telescopes such as SPTPol \cite{SPTPol}
and ACTPol \cite{ACTPol}, we work with patches of the sky
for which the ``flat sky approximation" (see e.g. \cite{flatsky})
is applicable. 

We find that the angular power spectrum of B-mode polarization
due to cosmic strings has a very similar shape as that 
of the gravitational lensing contribution. For values of $G \mu$
comparable to the current upper bound, the amplitude
of the power spectrum is smaller than the lensing signal
caused by the Gaussian perturbations. On the other hand,
the amplitude is sufficiently large such that the string signals
(which are highly non-Gaussian) ought to be visible in
position space analyses on B-mode maps.

\section{Polarization Signal of a Cosmic String Wake}

Wakes arise as a consequence of the geometry of space
perpendicular to a long straight string. Space is locally
flat, but globally it corresponds to a cone with deficit
angle \cite{deficit}
\be
\alpha \, = \, 8 \pi G \mu  \, ,
\ee
with the tip of the cone coinciding with the location
of the string.

On scales comparable to the Hubble length $t$, the long
string network is curved, which induces relativisitic transverse
velocities of the string.  A moving string yields a velocity
kick towards the plane in the wake of the string, which in
turn leads to a region behind the string with twice the
background density. The length of the wake is set by
the length of the cosmic string segment, i.e. by the
curvature radius of the long string, and is $c_1 t_i$ for
a string passing through the gas at time $t_i$. Here, $c_1$
is a numerical constant of order $1$. The depth of the
wake is determined by the string velocity $v_s$ and is
$v_s \gamma_s t_i$, where $\gamma_s$ is the relativistic
gamma factor associated with the velocity $v_s$.
The mean initial thickness of the string wake is
$4 \pi G \mu v_s \gamma_s t_i$.

The planar dimensions of the wake are fixed in comoving
coordinates. The wake thickness, on the other hand,
increases due to gravitational accretion from above and
below onto the string wake \footnote{Note that once
formed, the wake will persist even after the string
segment which has seeded it is no longer present.}. 
This accretion process can
be studied \cite{wakeZel} by means of the Zel'dovich
approximation \cite{Zel} with the result that the comoving
distance $q_{nl}(t, t_i)$ from the central plane of the wake which is
beginning to fall in at time $t > t_i$ is given by \cite{BDH}
\be
q_{nl}(t, t_i) \, = \, 
\frac{24 \pi}{5} G \mu v_s \gamma_s (z(t_i) + 1)^{-1/2} t_0 \bigl( \frac{t}{t_i} \bigr)^{2/3} \, ,
\ee
where $z(t)$ is the cosmological redshift at time $t$, and $t_0$ is
the present time. The last factor corresponds to
the growth factor from linear cosmological perturbation theory.
Note that in the case of wakes formed before the time $t_{eq}$
of equal matter and radiation this formula needs to be modified in two
ways: firstly the linear growth factor must be replaced by
$\frac{t}{t_{eq}}$ since the accretion of cold dark matter starts
only at $t_{eq}$. Secondly, the formula
\be
(z(t) + 1) t \, = \, (z(t) + 1)^{- 1/2} t_0
\ee
must be modified to read
\be
(z(t) + 1) t \, = \, (z(t) + 1)^{- 1} (z(t_{eq}) + 1)^{1/2} t_0 \, .
\ee
To summarize this discussion: the comoving size to which a wake formed
at time $t_i$ grows by time $t$ is
\be  \label{size}
 c_1 {t^c}_i  ~ \times~  v_s \gamma_s {t^c}_i ~ \times ~ 
\frac{24 \pi}{5} G \mu v_s \gamma_s {t^c}_i \bigl( \frac{t}{t_i} \bigr)^{2/3} \, ,
\ee
where $t^c$ indicates the comoving distance corresponding to $t$:
\be
t^c \, = \,  (z(t) + 1) t \, ,
\ee
and where, as mentioned above, for $t_i < t_{eq}$ the gravitational growth
factor involves the time $t_{eq}$ instead of the time $t_i$.

Since the string wake is a region of enhanced free electrons, CMB photons
emitted at the time of recombination acquire extra polarization when they pass
through a wake. Since the quadrupole direction is statistically uncorrelated
with the tangent vector of the string and its velocity vector, statistically
an equal strength E-mode and B-mode signal is produced.

In general, polarization can be described by a magnitude $P$ and a
direction $\alpha$. In terms of the Stokes parameters $Q$ and $U$, the
amplitude $P$ and angle $\alpha$ are given by
\bea \label{angle}
P \, &=& \, \sqrt{Q^2 + U^2} , \, \nonumber \\
\alpha \, &=& \, \frac{1}{2} {\rm{arctan}} \bigl( \frac{U}{Q} \bigr) \, .
\eea
The Stokes parameters in turn determine the $E$ and $B$ modes of
the polarization.

The strength of the polarization signal is determined by the CMB
quadrupole $Q_{quad}$, the column density
of free electrons which the CMB photons cross when passing through the
wake, and by the Thompson cross section $\sigma_T$. The column 
density of free electrons is given by the ionization fraction $f(t)$ of the wake
multiplied by the baryon column density, which in turn is determined
by the baryon fraction $\Omega_B$, the proton mass $m_p$ 
and the wake thickness. The polarization
signal of a single wake segment has a characteristic pattern which is
depicted in Fig. 1. The polarization strength is smallest 
at the position of the string (since this is the tip of the wake wedge),
and it is largest at the trailing edge. The polarization direction is roughly
constant across the wedge as long as wakes are considered which cover
a small fraction of the sky (and we will see that it is small wakes which
dominate the angular power spectrum). We see that the wake produced
by a string segment will lead to a rectangular region in the sky with extra
polarization.

\begin{figure} 
\includegraphics[height=6cm]{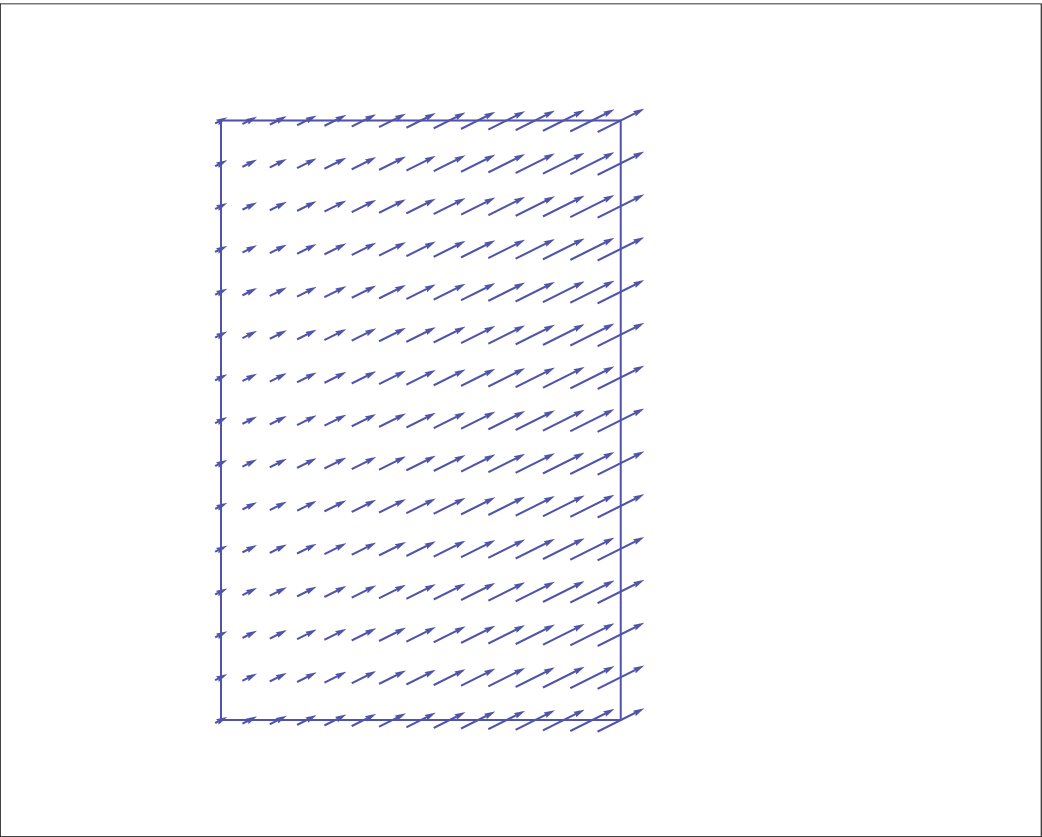}
\caption{Sketch of the polarization signal on the sky of a single string segment.
The amplitude of the polarization at any point on the rectangle which experiences
extra CMB polarization due to the string wake is indicated by the length of the
arrow, the direction of polarization is indicated by the arrow.}    
\label{wakevisual}
\end{figure}

The mean strength of the polarization signal $P$ of a wake produced
at time $t_i$ and intersected by our past light cone at time $t$,
averaged across the polarization rectangle in the sky, is \cite{BDH}
\bea \label{result}
\frac{P}{Q_{quad}} \, &\simeq& \, \frac{24 \pi}{25}  \bigl( \frac{3}{4 \pi} \bigr)^{1/2} \sigma_T f(t) G \mu
v_s \gamma_s \nonumber \\
& & \times \Omega_B \rho_c(t_0) m_p^{-1} t_0 \bigl( z(t) + 1 \bigr)^2 \bigl( z(t_i) + 1 \bigr)^{1/2} \, .
\eea
where $\rho_c(t_0)$ is the energy density of a spatially flat universe today. As
we will see in the following section, the wakes which contribute the
largest amount to the angular B-mode polarization power spectrum are those
created at around the time $t_{eq}$ of equal matter and radiation for
which our past light cone intersects the wake at a time $t$ just after
recombination. Hence, to get a feeling for the order of magnitude of
the polarization signal of a string wake we insert the numerical
values for the dimensional constants $\rho_c(t_0), m_p, \sigma_T$
and $t_0$ and normalize the redshifts at a value of $10^3$ to
obtain \footnote{Note the typo in the second to last factor of Eq. (19)
in \cite{BDH}.}
\be
\frac{P}{Q_{quad}}(t, t_i) \, \sim \, f(t) G \mu v_s \gamma_s \Omega_B 
\bigl( \frac{z(t) + 1}{10^3} \bigr)^2 \bigl( \frac{z(t_i) + 1}{10^3} \bigr)^{1/2}
10^7 \, .
\ee

For a value of $G \mu = 10^{-7}$ comparable to the current upper
bound on the string tension, then we find for $t_i \simeq t \simeq t_{rec}$
\be \label{resulteval}
\frac{P}{Q_{quad}} \, \sim \, \Omega_B \, ,
\ee
where we have inserted the value $v_s \gamma_s \simeq 1$ and used
the fact that just after recombination $f \sim 1$. We thus see that in
position space, the polarization signal of a cosmic string wake
has a large amplitude. 

The distribution of cosmic strings is highly non-Gaussian. More
specifically, string wakes only cover a fraction of about $G \mu$
of the volume of space at $t_{eq}$. Hence, when computing the
power spectrum we expect the non-Gaussianness of the string
distribution to lead to an amplitude of the power spectrum of
$P/Q_{quad}$ which is suppressed by this small factor compared to unity,
and thus we expect that the string signal is much harder to
identify in the power spectrum.

We end this section with a discussion of the formula for polarization
which will underly our study. The starting point is
the fact that the linear polarization of the CMB is described by
two Stokes parameters $Q({\bf x})$ and $U({\bf x})$, where
${\bf x}$ are Cartesian coordinates on a patch of 
the celestial sphere. The Stokes parameters
determine the E and B mode polarization functions. In the
flat sky approximation, these are given by \cite{Bunn}
\bea \label{basic}
Q({\bf x}) \, & = &  \, \int \frac{d^2 {\bf l}}{(2 \pi)^2} 
\bigl[ E({\bf l}) {\rm cos}(2 \phi_l) - B({\bf l}) {\rm sin}(2 \phi_l) \bigr]
e^{i {\bf l} \cdot {\bf x}} \nonumber \\
U({\bf x}) \, & = & \, \int \frac{d^2 {\bf l}}{(2 \pi)^2}
\bigl[ E({\bf l}) {\rm sin}(2 \phi_l) + B({\bf l}) {\rm cos}(2 \phi_l) \bigr]
e^{i {\bf l} \cdot {\bf x}}  \,  \nonumber \\
& &\, ,
\eea
where ${\bf l} = l ({\rm cos}(\phi_l), {\rm sin}(\phi_l))$.
These formulas can be inverted to give $E$ and $B$ as functions
of $U$ and $Q$, which then are determined in turn by 
the total amplitude $P$ and the angle $\alpha$ via (\ref{angle}).
In the flat sky Fourier space the inverted formulas are
\bea \label{moderesult}
E({\bf l}) \, &=& \, {\tilde Q}({\bf{l}}) {\rm cos}(2  \phi_l)  + {\tilde U}({\bf{l}}) {\rm sin} (2 \phi_l) \nonumber \\
B({\bf l}) \, &=& \, - {\tilde Q}({\bf{l}}) {\rm sin}(2  \phi_l)  + {\tilde U}({\bf{l}}) {\rm cos} (2 \phi_l) \, ,
\eea
where ${\tilde Q}$ and ${\tilde U}$ are the Fourier coefficients of $Q$ and $U$, respectively.

In the following section we use the above basic equations
(\ref{basic}) to compute the angular power spectrum of
both E-mode and B-mode polarization, starting with
the expression (\ref{result}) for the polarization of
a single wake, and summing over all wakes which are
crossed by the past light cone.

\section{Computation of the Angular Power Spectrum}

We have performed two independent numerical computations
of the angular power spectrum of CMB polarization produced
by a scaling distribution of cosmic string wakes \cite{Nick, Grant}. Since
good angular resolution is more important to identify
cosmic string signals than full sky coverage, we have
applications to experiments such as the South Pole
Telescope \cite{SPT} and the Atacama Cosmology Telescope \cite{ACT}
in mind which map out portions of the sky to which the
flat sky approximation can be applied. Hence, in both
of our computations we make use of this approximation.

Our starting point is the formula (\ref{result}) for
the amplitude of polarization from a single string wake
created at time $t_i$ which is being crossed by our past
light cone at time $t > t_i$. The angular power spectra
of E and B mode polarization obtain contributions from
each string wake which is crossed by the past light cone.
Since the centers and orientations of the string wakes
are uncorrelated, the contributions of all strings are
independent, and hence the power spectra are obtained
by integrating the power spectra produced by a single
string wake over the times $t_i$ and $t$. 

According to the string scaling solution, there are
a fixed number $n_w$ of string segments per Hubble volume.
They live for a Hubble expansion time. Thus, to take
care of the integration over the formation time $t_i$,
we divide the time interval from $t_{eq}$ to the present
time into Hubble expansion times, pick the initial value
$t_i$ in each interval, place a number $N$ of string
wakes at random in each Hubble volume at time $t_i$,
and sum over $t_i$. The reason for only considering
formation times after $t_{eq}$ is the fact that earlier
strings are smaller but start to accrete cold dark matter
only at the time $t_{eq}$.

For each time $t_i$, we have to determine how many string
wakes formed at that time are crossed by the past light
cone at time $t$. We again break up the time interval
for $t$ into Hubble time steps, and compute the
number of wakes formed in the k'th Hubble time step
centered at $t_k$ crossing the past light cone in the
Hubble time step centered at the discrete values of $t$.
The formula (\ref{result}) then
can be used to compute the contribution of each such string
wake to the power spectra. Finally, we must integrate
over $t > t_j$. Obviously the condition $t > t_{rec}$ must be 
imposed (since the polarization we consider is produced
by the decoupled CMB radiation being scattered in the wake).  

Based on the statistical independence of the positions and
orientations of string wakes, the angular power
spectrum of a scaling solution of strings is
\be \label{finalpower}
C_l \, = \, \sum_{t_i}  \sum_{t > t_i} N_w(t, t_i) C_l(t, t_i) \, ,
\ee
where $N_w(t, t_i)$ is the number of wakes laid down at time
$t_i$ (within one Hubble expansion time) which intersect
the past light cone in a Hubble expansion time about time $t$,
and $C_l(t, t_i)$ is the angular power spectrum of a string
wake produced at time $t_i$ which intersects the past light
cone at time $t$.

The number of string wakes created in a Hubble expansion
time around $t_i$ and intersecting the past light cone of the
observing area in a Hubble time about $t$ is obtained by
computing how many comoving Hubble volumes at $t_i$
are contained in the comoving past light cone (PLC) of the
observing volume at time $t$, multiplied by the number
$n_w$ of strings per Hubble volume:
\be \label{number}
N_w(t, t_i) \, = \, n_w \frac{{\rm Vol(PLC)}_{\rm com}(t)}{{\rm Vol(Hubble)}_{\rm com}(t_i)} \, ,
\ee
where the notation is explained in the text above.

The  comoving Hubble volume at time $t_i$ is given by
\be
{\rm Vol(Hubble)}_{\rm com}(t_i) \, = \, \frac{9 \pi}{2} t_0^2 t_i \, ,
\ee
and the volume of the past light cone of the observation area of
angular sizes $\theta_1$ and $\theta_2$ is
\be
{\rm Vol(PLC)}_{\rm com}(t) \, = \, 18 \theta_1 \theta_2 t_0^3 \bigl( 1 - a(t)^{1/2} \bigr)^2 \, a(t)^{1/2} ,
\ee
where we have normalized the scale factor to be one at the present
time $t_0$ \footnote{This result is obtained by computing the comoving
distance light travels from $t$ to $t_0$, taking the square of this
quantity, multiplying with the comoving Hubble time at $t$ and finally
multiplying by $\theta_1 \times \theta_2$.}. 
Inserting these results into (\ref{number}) we get
\be \label{wakenumber}
N_w(t, t_i) \, = \, n_w \frac{4}{\pi} \theta_1 \theta_2  \bigl( \frac{t_0}{t_i} \bigr)
\bigl( \frac{t}{t_0} \bigr)^{1/3} \bigl( 1 - (\frac{t}{t_0} )^{1/3} \bigr)^2 \, .
\ee

Next we need to compute the angular power spectrum $C_l(t, t_i)$ of
a string wake created at time $t$ intersecting the past light cone
at time $t$. Since the angular power spectrum involves an average over
space and an average over angles we can - without loss of generality - set
the center of the wake to be at the origin of the flat sky coordinates, and
take the string which creates the wake to be moving along the x-axis. Thus,
the polarization amplitude is a linearly increasing function of $x$ and is
independent of $y$, and its average magnitude $P(t, t_i)$ is given by (\ref{result}):
\be
P(x, y, t, t_i) \, = \, P(t, t_i) \Theta(|x| < \frac{d}{2}) \Theta(|y| < \frac{w}{2}) \bigl( \frac{2}{d} x + 1 \bigr) \, ,
\ee
where $d = d(t, t_i)$ and $w = w(t, t_i)$ give the angular dimensions of the 
wake seen in the sky. These angular dimensions are given by
\bea
d(t, t_i) \, &=& \, v_s \gamma_s \omega(t, t_i) \, \nonumber \\
w(t, t_i) \, &=& \, c_1 \omega(t, t_i) \, ,
\eea
where, as discussed in Section 2, $v_s$ is the velocity of the string, $\gamma_s$
is the associated relativistic gamma factor, and $c_1$ is a constant of order one
describing the coherence length of the string network in units of $t$. The
quantity $\omega(t, t_i)$ is the angular scale of the comoving Hubble radius
at $t_i$ as seen at time $t > t_i$. It is given by
\be
\omega(t, t_i) \, = \, \frac{ (t_i / t_o)^{1/3}}{2 ( 1 - (t / t_0)^{1/3})} \, .
\ee

The magnitude ${\tilde{P}}({\bf{l}})$ of the polarization of the string wake is
given by the flat space Fourier integral of the position space polarization
$P(x, y, t)$. This integral can be performed exactly analytically, with the
result
\bea \label{result3}
& & {\tilde{P}}({\bf{l}}) \, = \, P(t, t_i) \times \\
& & \, \, \frac{4 {\rm{sin}}(l_y w / 2) 
\bigl( i l_x d {\rm{cos}}(l_x d / 2) + (- 2i + l_x d) {\rm{sin}}(l_x d / 2) \bigr)}{l_x^2 l_y d} \, ,
\nonumber
\eea
where $P(t, t_i)$ is given by (\ref{result}).

To evaluate the above result it is necessary to make use of the ionization
fraction $f(t)$. We have used an analytical fit to the time dependence of
$f(t)$ determined in \cite{Gil}. At the time of recombination (corresponding
to a redshift of $z = 10^3$), the ionization fraction is unity. By a redshift of
$z = 500$ the fraction has decreased to $f(t) \sim 10^{-3}$. This effect
increases the importance of wakes crossing the Hubble radius within the
first Hubble expansion time after $t_{rec}$ relative to those crossing
at later times. After reionization $f(t)$ jumps back up to close to unity.

In the flat sky approximation, the angular power spectrum $C_l^{XX}$
of the quantity $X$ (where $X$ can stand for temperature $T$ or polarizations
$E$ and $B$) is given by
\be \label{corfct}
C_l^{XX} \, = \, \langle X({\bf{l}})X^*( {\bf{l}}) \rangle \, ,
\ee
where $l$ is the magnitude of ${\bf{l}}$ and the angular brackets indicate
averaging over angles of ${\bf{l}}$ and over the random variables describing
the position of the wake and the orientation with respect to the CMB quadrupole,
the angle called $\alpha$ earlier in the text \footnote{Note that the definition of
the angular correlation functions involves averaging over the entire sky. In the
case of surveys and theoretical simulations involving only a fraction of the sky,
the result obtained by integrating over the solid angle $\Omega$ for which
data is present must be multiplied by $4 \pi / \Omega$. In our case, $\Omega$
is given by  $\Omega = \theta_1 \theta_2$. As a consequence, the 
factor $\theta_1 \theta_2$ in 
(\ref{wakenumber}) is replaced by $4 \pi$.}. The cross-correlation functions
are described by a similar formula
\be \label{corfct2}
C_l^{XY} \, = \, \langle X({\bf{l}})Y^*( {\bf{l}}) \rangle \, ,
\ee

First we show that the angular correlation functions of E and B mode polarization
are the same
\be
C_l^{BB} \, = \, C_l^{EE} \, ,
\ee
and that the cross-correlation function between E and B mode polarization
vanishes
\be
C_l^{EB} \, = \, 0 \, .
\ee
To see this, we make use of the random orientation of the string motion
relative to the CMB quadrupole to write
\bea \label{ansatz2}
Q \, &=& \, P {\rm{cos}}(2 \alpha) \, \nonumber \\
U \, &=& \, P {\rm{sin}}(2 \alpha) \, ,
\eea
where $\alpha$ is a random angle. Inserting (\ref{ansatz2})
into (\ref{moderesult}) and the result of that into the
expressions (\ref{corfct}) and (\ref{corfct2}) we
find that upon averaging over $\alpha$ the expressions
for the E and B mode polarization power spectra are
the same, and that the cross-correlation function vanishes.
More specifically, we obtain the following expression for
the contribution of a string wake created at time $t_i$
and intersecting the past light cone at time $t$
\be \label{singlepower}
C_l^{EE}(t, t_i) \, = \, C_l^{BB}(t, t_i) \, = \,  \frac{1}{4 \pi} \int_0^{2 \pi} |{\tilde{P}}(t, t_i)|^2 d \phi_l \, ,
\ee
where ${\tilde{P}}(t, t_i)$ is given by (\ref{result3}).

The final power spectrum of B and E mode polarization is obtained
by inserting (\ref{singlepower}) and (\ref{wakenumber}) into 
(\ref{finalpower}) and performing the double sum. It is
straightforward to see that (even before taking into account
the time dependence of $f(t)$) the sum is dominated by the earliest
times $t_i$ and by the earliest times $t$ consistent with $t > t_{rec}$.
The time dependence of $f(t)$ further increases the importance
of wakes with early $t$ relative to those with later $t$.
The sum in (\ref{finalpower}) was computed numerically, taking
into account the time dependence of $f(t)$.
The results are shown in Figures 2 and 3. Figure 2 shows
$\sqrt{l(l+1)C_l/(2\pi)}$, Figure 3 $C_l$ alone (showing
the large angle plateau).

\begin{widetext}

\begin{figure} 
\includegraphics[height=6cm]{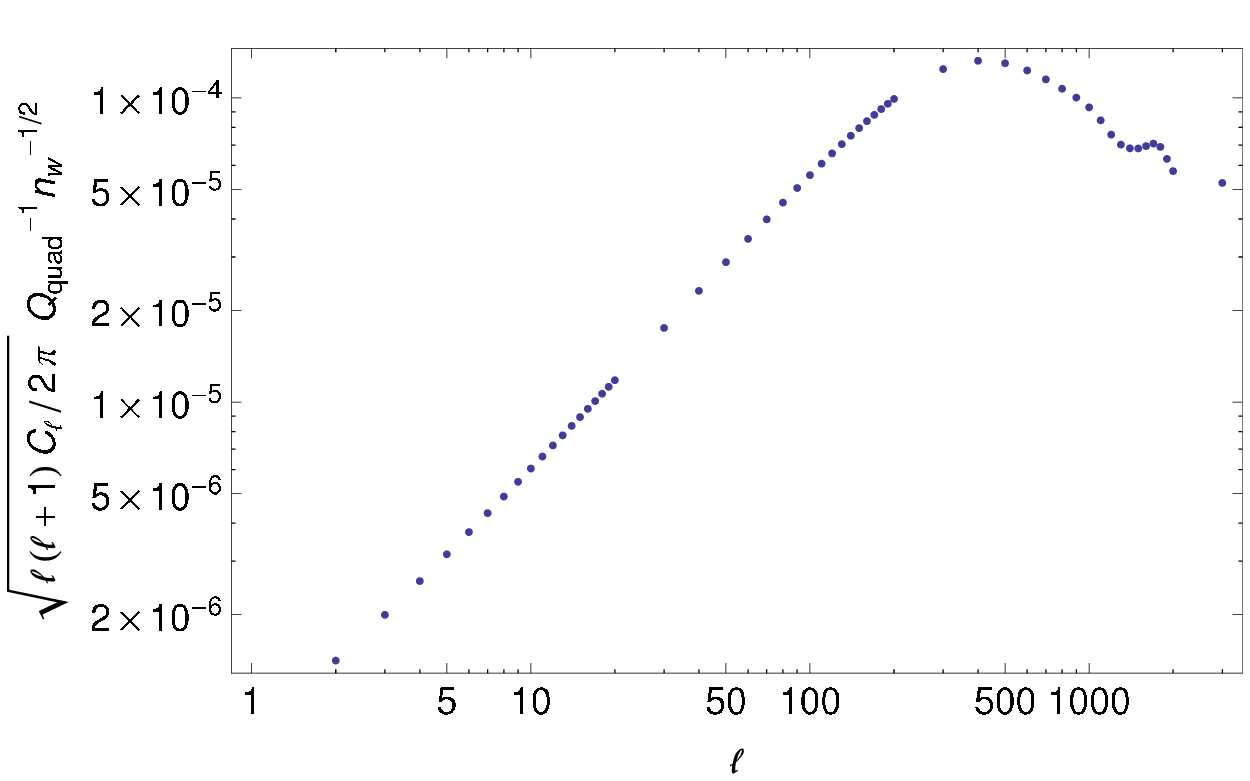}
\caption{The angular power spectrum $\sqrt{l(l + 1) C_l / 2 \pi}$
of B-mode polarization (in units of $Q_{quad}$)
for a scaling solution of cosmic strings, for a value of $G \mu = 10^{-7}$,
for $c_1 = 1$, $v_s \gamma_s = 1$.  The vertical axis
is $\sqrt{l (l + 1) C_l / (2 \pi)}$, the horizontal axis is $l$. Both axes are
logarithmic. We have normalized to the number $n_w$ of wakes
per Hubble volume.}
\label{angular}
\end{figure}

\end{widetext}

\begin{figure} 
\includegraphics[height=5cm]{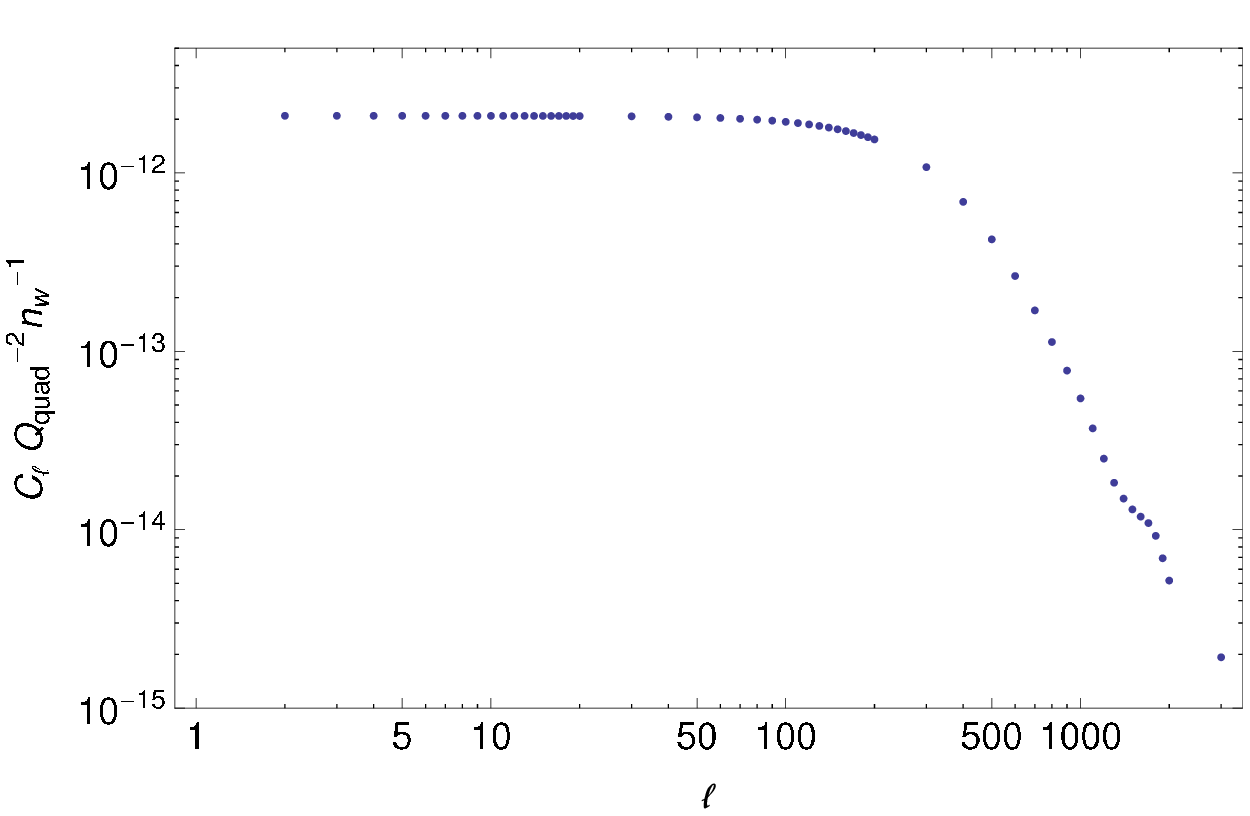}
\caption{The angular power spectrum $C_l$ of B-mode polarization
for a scaling solution of cosmic strings, for a value of $G \mu = 10^{-7}$,
for $c_1 = 1$, $v_s \gamma_s = 1$.  The vertical axis
is $C_l / Q^2$, the horizontal axis is $l$. Both axes are
logarithmic.}
\label{angular2}
\end{figure}

The amplitude of the angular power spectrum can be estimated
analytically. Combining the formulas for the ingredients in
(\ref{finalpower}) and restricting the sum to the dominant wakes,
those with $t_i = t_{eq}$ and $t = t_{rec}$, we find
\be
C_l^{BB} \, \sim \, N z_{eq}^{-2} P^2(t_{eq}, t_{rec}) \, .
\ee
Inserting the value of the local polarization of the dominant
wakes from (\ref{resulteval}) for the value $G \mu = 10^{-7}$
we obtain an amplitude of the order $10^{-11} Q_{quad}^2$ \footnote{We have
used the value $\Omega_B = 0.022$.} which agrees
well with the numerical results. Note that the amplitude of
$\sqrt{l(l+1) C_l}$ is linear in $G \mu$.

Since it is wakes created at time $t_{eq}$ intersecting the past light cone 
at time $t = t_{rec}$ which dominate the angular correlation function,
the position of the peak in the angular power spectrum in Figure 2
corresponds to the angular size of these dominant wakes. From
the analytical formula (\ref{result3}), it follows that in the low $l$
limit, the amplitude of $C_l$ is constant. This explains the slope of
the power spectrum of $C_l$ at small values of $l$. The decrease
in the angular power spectrum for values of $l$ larger than that
corresponding to the peak position can be argued for from the
Riemann-Lebesgue lemma. However, the shape in this high $l$
region depends on the assumption that wakes have no small-scale
structure, and that wakes created before $t_{eq}$ have a
negligible effect.

In Figure 4 we show the contributions of different Hubble time
intervals of $t_i$ (and for the dominant value of $t$ which corresponds to 
$t = t_i$) to the angular power spectrum. The integer $n$ labels the Hubble
expansion time step, with $n = 1$ being the time step immediately
after the time of equal matter and radiation. The figure
shows that the contribution of wakes decreases as $t_i$ increases.
It also shows that the peak position shifts to larger angular
scales as $t_i$ increases since the wake size increases. Note that
the increase in the amplitude between $n = 10$ and $n = 14$ is
due to the increase of the ionization fraction $f$ for times $t$
after reionization.

\begin{widetext}

\begin{figure} 
\includegraphics[height=6cm]{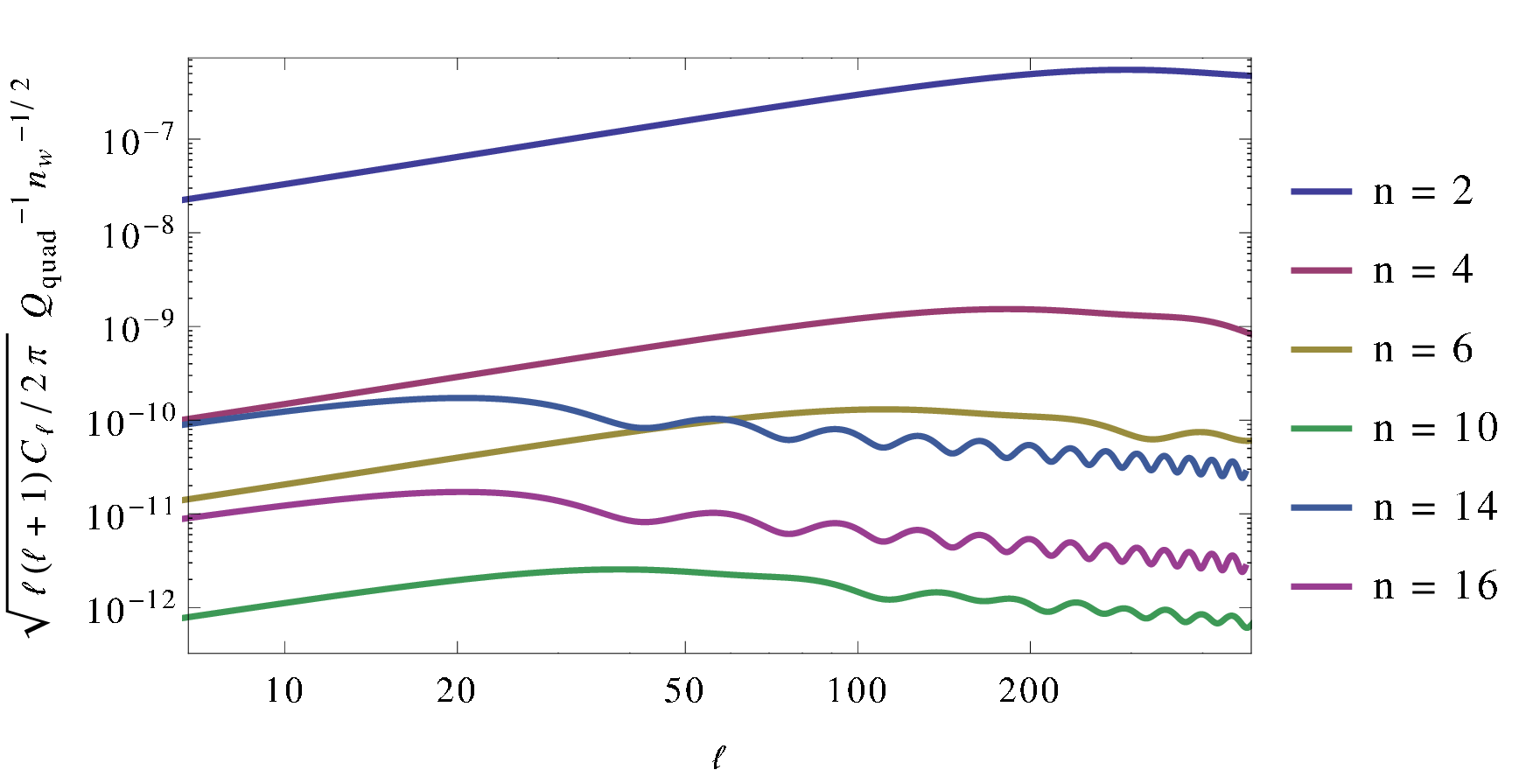}
\caption{The contribution of wakes for different Hubble
time steps in $t_i$ (and $t = t_i$) to the angular power spectrum 
of B mode polarization. The axes and string parameter values are the
same as in the previous figure. The integer $n$ labels the Hubble
expansion step, with $n = 1$ corresponding to the first Hubble
time step after the time of equal matter and radiation.}
\label{angularsteps}
\end{figure}

\end{widetext}

\section{Discussion and Conclusions}

We have computed the power spectrum of E and B mode polarization
produced by a scaling distribution of cosmic strings. As
already realized in \cite{BDH}, cosmic strings produce
B mode polarization at leading order. 

We find that the contribution to the E and B mode power
spectra is dominated by the earliest wakes (those created
at around $t_{eq}$, and those which are intersected by
our past light cone closest to the time $t_{rec}$ of
last scattering. For values of the string tension
$G \mu = 10^{-7}$ close to the current observational
bound, the predicted polarization signal in position
space has an amplitude $P/Q_{quad} \sim \Omega_B$ \cite{BDH}.
Cosmic string wakes, however, correspond to a very
non-Gaussian distribution of density enhancements.
Hence, the string signal in the power spectrum is
greatly suppressed. For $G \mu = 10^{-7}$ we find
an amplitude of the power spectrum $\sqrt{l(l+1)C_l}$ 
of $P/Q_{quad}$ which is (for values of $l$ smaller than
that corresponding to the peak position) of order 
$ l z_{eq}^{-3/2} \Omega_B$, much smaller than the 
position space signal, but only one order of magnitude
smaller than the predicted B-mode signal from gravitational lensing
produced by the dominant Gaussian fluctuations. A rough
way of understanding the suppression of the signal
in the power spectrum compared to the signal in
position space is to realize that the string wakes
which dominated the polarization signal occupy
a small fraction of space, and that hence
a suppression by this factor is to be expected.
The magnitude of the polarization signal scales
linearly in $G \mu$.

Note that the power spectrum of B mode polarization
from string wakes has the same shape as that of the
lensing signal. The peak position is located at
a value of $l \sim 500$ which corresponds to the
angular scale of the comoving Hubble radius at the
time $t_{eq}$ of equal matter and radiation. The
slope of $\sqrt{l(l+1)C_l}$ for small $l$ can be explained in terms of
Poisson superposition of the effect of the dominant
wakes, in the same way that the slope of the lensing
signal on large angular scales can be understood
via the Poisson distribution of the small but
dominant lensing kicks \footnote{We thank Gil Holder
for discussions on this point.}. The peak position
of the lensing signal is related to the scale
where the matter power spectrum turns over and hence
is comparable to the peak position of the string signal.

The fact that the amplitude of the angular power spectrum
of string-induced B-mode polarization is (for a value
of $G \mu = 10^{-7}$) an order of magnitude
smaller than that of the B-mode
polarization induced by lensing will make it hard to
see the string-induced signal in Fourier space. However,
the fact that the position space signal has a specific
geometry will make it easy to detect the string signal
above the lensing noise using a position space
analysis of the lensing maps.
 
We conclude that searches for cosmic strings in B-mode
polarization must be done in position space. We
must search for the distinctive geometrical patterns
on the sky which string wakes predict. For example,
one could use edge detection algorithms like the
Canny algorithm \cite{Canny} to search for the
distinctive edges in the B mode sky produced by
string wakes, in a similar way that this algorithm
was used to search for cosmic string wake signals in
CMB temperature maps \cite{Danos}. The studies of
\cite{Danos} in fact showed that the string signals
can be dug out of a Gaussian noise with a much larger
amplitude. In current work, we are studying the
prospects for the application of the Canny algorithm to
polarization maps.

\begin{acknowledgments} 
 
This work is supported in part by a NSERC Discovery Grant (RB), 
by funds from the CRC Program (RB), and by both an NSERC Canada
Graduate Scholarship and a McGill Chalk-Rowles Fellowship to GS. 
We wish to thank Matt Dobbs, Duncan Hanson, Gil Holder and Elaine Roebbers
for useful discussions.

\end{acknowledgments}


\begin{thebibliography}{99} 

\bibitem{BDH}
R.~J.~Danos, R.~H.~Brandenberger and G.~Holder,
  ``A Signature of Cosmic Strings Wakes in the CMB Polarization,''
  Phys.\ Rev.\ D {\bf 82}, 023513 (2010)
  [arXiv:1003.0905 [astro-ph.CO]].
  %%CITATION = ARXIV:1003.0905;%%

\bibitem{White}
W.~Hu and M.~J.~White,
  ``A CMB polarization primer,''
  New Astron.\  {\bf 2}, 323 (1997)
  [astro-ph/9706147].
  %%CITATION = ASTRO-PH/9706147;%%

\bibitem{Emodeobs}
J.~Kovac, E.~M.~Leitch, CPryke, J.~E.~Carlstrom, N.~W.~Halverson and W.~L.~Holzapfel,
  ``Detection of polarization in the cosmic microwave background using DASI,''
  Nature {\bf 420}, 772 (2002)
  [astro-ph/0209478].
  %%CITATION = ASTRO-PH/0209478;%%

\bibitem{Future}
D.~Baumann {\it et al.}  [CMBPol Study Team Collaboration],
  ``CMBPol Mission Concept Study: Probing Inflation with CMB Polarization,''
  AIP Conf.\ Proc.\  {\bf 1141}, 10 (2009)
  [arXiv:0811.3919 [astro-ph]].
  %%CITATION = ARXIV:0811.3919;%%

\bibitem{Hanson}
D.~Hanson {\it et al.}  [SPTpol Collaboration],
  ``Detection of B-mode Polarization in the Cosmic Microwave Background with Data from the South Pole Telescope,''
  arXiv:1307.5830 [astro-ph.CO].
  %%CITATION = ARXIV:1307.5830;%%
  
\bibitem{lensing}
M.~Zaldarriaga and U.~Seljak,
  ``Gravitational lensing effect on cosmic microwave background polarization,''
  Phys.\ Rev.\ D {\bf 58}, 023003 (1998)
  [astro-ph/9803150].
  %%CITATION = ASTRO-PH/9803150;%%

\bibitem{Starob}
A.~A.~Starobinsky,
  ``Relict Gravitation Radiation Spectrum and Initial State of the Universe. (In Russian),''
  JETP Lett.\  {\bf 30}, 682 (1979)
  [Pisma Zh.\ Eksp.\ Teor.\ Fiz.\  {\bf 30}, 719 (1979)].
  %%CITATION = JTPLA,30,682;%%

\bibitem{holy}
R.~H.~Brandenberger,
  ``Is the Spectrum of Gravitational Waves the 'Holy Grail' of Inflation?,''
  arXiv:1104.3581 [astro-ph.CO].
  %%CITATION = ARXIV:1104.3581;%%

\bibitem{ShellVil}
A. Vilenkin and E.P.S. Shellard, \textit{Cosmic Strings and other
Topological Defects} (Cambridge Univ. Press, Cambridge, 1994).

\bibitem{HK}
M.~B.~Hindmarsh and T.~W.~B.~Kibble,
  ``Cosmic strings,''
  Rept.\ Prog.\ Phys.\  {\bf 58}, 477 (1995)
  [arXiv:hep-ph/9411342].
  %%CITATION = HEP-PH 9411342;%%

\bibitem{RHBrev}
R.~H.~Brandenberger,
  ``Topological defects and structure formation,''
  Int.\ J.\ Mod.\ Phys.\ A {\bf 9}, 2117 (1994)
  [arXiv:astro-ph/9310041].
  %%CITATION = ASTRO-PH 9310041;%%

\bibitem{Kibble}
 T.~W.~B.~Kibble,
  ``Phase Transitions In The Early Universe,''
  Acta Phys.\ Polon.\  B {\bf 13}, 723 (1982);\\
  %%CITATION = APPOA,B13,723;%%
  T.~W.~B.~Kibble,
  ``Some Implications Of A Cosmological Phase Transition,''
  Phys.\ Rept.\  {\bf 67}, 183 (1980).
  %%CITATION = PRPLC,67,183;%%

\bibitem{Dvorkin}
 C.~Dvorkin, M.~Wyman and W.~Hu,
  ``Cosmic String constraints from WMAP and SPT,''
  arXiv:1109.4947 [astro-ph.CO].
  %%CITATION = ARXIV:1109.4947;%%

\bibitem{PlanckCS}
P.~A.~R.~Ade {\it et al.}  [Planck Collaboration],
  ``Planck 2013 results. XXV. Searches for cosmic strings and other topological defects,''
  arXiv:1303.5085 [astro-ph.CO].
  %%CITATION = ARXIV:1303.5085;%%

\bibitem{previous}
L.~Pogosian, S.~H.~H.~Tye, I.~Wasserman and M.~Wyman,
  ``Observational constraints on cosmic string production during brane
  inflation,''
  Phys.\ Rev.\  D {\bf 68}, 023506 (2003)
  [Erratum-ibid.\  D {\bf 73}, 089904 (2006)]
  [arXiv:hep-th/0304188];\\
  %%CITATION = PHRVA,D68,023506;%%
M.~Wyman, L.~Pogosian and I.~Wasserman,
  ``Bounds on cosmic strings from WMAP and SDSS,''
  Phys.\ Rev.\  D {\bf 72}, 023513 (2005)
  [Erratum-ibid.\  D {\bf 73}, 089905 (2006)]
  [arXiv:astro-ph/0503364];\\
  %%CITATION = PHRVA,D72,023513;%%
A.~A.~Fraisse,
  ``Limits on Defects Formation and Hybrid Inflationary Models with
  Three-Year WMAP Observations,''
  JCAP {\bf 0703}, 008 (2007)
  [arXiv:astro-ph/0603589];\\
  %%CITATION = JCAPA,0703,008;%%
U.~Seljak, A.~Slosar and P.~McDonald,
  ``Cosmological parameters from combining the Lyman-alpha forest with CMB,
  galaxy clustering and SN constraints,''
  JCAP {\bf 0610}, 014 (2006)
  [arXiv:astro-ph/0604335];\\
  %%CITATION = JCAPA,0610,014;%%
  R.~A.~Battye, B.~Garbrecht and A.~Moss,
  ``Constraints on supersymmetric models of hybrid inflation,''
  JCAP {\bf 0609}, 007 (2006)
  [arXiv:astro-ph/0607339];\\
  %%CITATION = JCAPA,0609,007;%%
R.~A.~Battye, B.~Garbrecht, A.~Moss and H.~Stoica,
  ``Constraints on Brane Inflation and Cosmic Strings,''
  JCAP {\bf 0801}, 020 (2008)
  [arXiv:0710.1541 [astro-ph]];\\
  %%CITATION = JCAPA,0801,020;%%
N.~Bevis, M.~Hindmarsh, M.~Kunz and J.~Urrestilla,
  ``CMB power spectrum contribution from cosmic strings using  field-evolution
  simulations of the Abelian Higgs model,''
  Phys.\ Rev.\  D {\bf 75}, 065015 (2007)
  [arXiv:astro-ph/0605018];\\
  %%CITATION = PHRVA,D75,065015;%%
N.~Bevis, M.~Hindmarsh, M.~Kunz and J.~Urrestilla,
  ``Fitting CMB data with cosmic strings and inflation,''
  arXiv:astro-ph/0702223;\\
  %%CITATION = ASTRO-PH/0702223;%%
R.~Battye and A.~Moss,
  ``Updated constraints on the cosmic string tension,''
  arXiv:1005.0479 [astro-ph.CO].
  %%CITATION = ARXIV:1005.0479;%%

\bibitem{Francis}
F.~Duplessis and R.~Brandenberger,
  ``Note on Structure Formation from Cosmic String Wakes,''
  JCAP {\bf 1304}, 045 (2013)
  [arXiv:1302.3467 [astro-ph.CO]].
  %%CITATION = ARXIV:1302.3467;%%

\bibitem{Oscar}
R.~H.~Brandenberger, R.~J.~Danos, O.~F.~Hernandez and G.~P.~Holder,
  ``The 21 cm Signature of Cosmic String Wakes,''
  JCAP {\bf 1012}, 028 (2010)
  [arXiv:1006.2514 [astro-ph.CO]].
  %%CITATION = ARXIV:1006.2514;%%

\bibitem{previous21cm}
R.~Khatri and B.~D.~Wandelt,
  ``Cosmic (super)string constraints from 21 cm radiation,''
  Phys.\ Rev.\ Lett.\  {\bf 100}, 091302 (2008)
  [arXiv:0801.4406 [astro-ph]];\\
  %%CITATION = PRLTA,100,091302;%%
 A.~Berndsen, L.~Pogosian and M.~Wyman,
  ``Correlations between 21 cm Radiation and the CMB from Active Sources,''
  arXiv:1003.2214 [astro-ph.CO].
  %%CITATION = ARXIV:1003.2214;%% 

\bibitem{CMBpolprevious}
U.~L.~Pen, U.~Seljak and N.~Turok,
  ``Power spectra in global defect theories of cosmic structure formation,''
  Phys.\ Rev.\ Lett.\  {\bf 79}, 1611 (1997)
  [arXiv:astro-ph/9704165].
%%CITATION = ASTRO-PH 9704165;%%  

\bibitem{RHBCSrev}
R.~H.~Brandenberger,
  ``Searching for Cosmic Strings in New Observational Windows,''
  arXiv:1301.2856 [astro-ph.CO].
  %%CITATION = ARXIV:1301.2856;%%

\bibitem{onescale}
L.~Perivolaropoulos,
  ``COBE versus cosmic strings: An Analytical model,''
  Phys.\ Lett.\  B {\bf 298}, 305 (1993)
  [arXiv:hep-ph/9208247];\\
  %%CITATION = PHLTA,B298,305;%%
L.~Perivolaropoulos,
  ``Statistics of microwave fluctuations induced by topological defects,''
  Phys.\ Rev.\  D {\bf 48}, 1530 (1993)
  [arXiv:hep-ph/9212228].
  %%CITATION = PHRVA,D48,1530;%%

\bibitem{wake}
J.~Silk and A.~Vilenkin,
  ``Cosmic Strings And Galaxy Formation,''
  Phys.\ Rev.\ Lett.\  {\bf 53}, 1700 (1984);\\
  %%CITATION = PRLTA,53,1700;%%
 M. Rees,
 ``Baryon concentrations in string wakes at $z \geq 200$:
 implications for galaxy formation and large-scale structure,"
 Mon. Not. R. astr. Soc. {\bf{222}}, 27p (1986);\\
T.~Vachaspati,
  ``Cosmic Strings and the Large-Scale Structure of the Universe,''
  Phys.\ Rev.\ Lett.\  {\bf 57}, 1655 (1986);\\
  %%CITATION = PRLTA,57,1655;%%
A.~Stebbins, S.~Veeraraghavan, R.~H.~Brandenberger, J.~Silk and N.~Turok,
  ``Cosmic String Wakes,''
  Astrophys.\ J.\  {\bf 322}, 1 (1987);\\
  %%CITATION = ASJOA,322,1;%%
J.~C.~Charlton,
  ``Cosmic String Wakes and Large Scale Structure,''
  Astrophys.\ J.\  {\bf 325}, 52 (1988);\\
%%CITATION = ASJOA,325,521;%%
T.~Hara and S.~Miyoshi,
  ``Formation of the First Systems in the Wakes of Moving Cosmic Strings,''
  Prog.\ Theor.\ Phys.\  {\bf 77}, 1152 (1987);\\
  %%CITATION = PTPKA,77,1152;%%
T.~Hara and S.~Miyoshi,
  ``Flareup of the Universe After Z Appproximately 10**2 for Cosmic String 
  Model,''
  Prog.\ Theor.\ Phys.\  {\bf 78}, 1081 (1987);\\
  %%CITATION = PTPKA,78,1081;%%
T.~Hara and S.~Miyoshi,
  ``Large Scale Structures and Streaming Velocities Due to Open Cosmic 
  Strings,''
  Prog.\ Theor.\ Phys.\  {\bf 81}, 1187 (1989).
  %%CITATION = PTPKA,81,1187;%%

\bibitem{Seljak}
U.~Seljak, U.~L.~Pen and N.~Turok,
  ``Polarization of the Microwave Background in Defect Models,''
  Phys.\ Rev.\ Lett.\  {\bf 79}, 1615 (1997)
  [arXiv:astro-ph/9704231];\\
  %%CITATION = PRLTA,79,1615;%%
 U.~Seljak and A.~Slosar,
  ``B polarization of cosmic microwave background as a tracer of strings,''
  Phys.\ Rev.\  D {\bf 74}, 063523 (2006)
  [arXiv:astro-ph/0604143];\\
  %%CITATION = PHRVA,D74,063523;%% 
L.~Pogosian, I.~Wasserman and M.~Wyman,
  ``On vector mode contribution to CMB temperature and polarization from  local
  strings,''
  arXiv:astro-ph/0604141;\\
  %%CITATION = ASTRO-PH/0604141;%%
L.~Pogosian and M.~Wyman,
  ``B-modes from Cosmic Strings,''
  Phys.\ Rev.\  D {\bf 77}, 083509 (2008)
  [arXiv:0711.0747 [astro-ph]];\\
  %%CITATION = PHRVA,D77,083509;%%
K.~Benabed and F.~Bernardeau,
  ``Cosmic string lens effects on CMB polarization patterns,''
  Phys.\ Rev.\  D {\bf 61}, 123510 (2000);\\
  %%CITATION = PHRVA,D61,123510;%%     
J.~Garcia-Bellido, R.~Durrer, E.~Fenu, D.~G.~Figueroa and M.~Kunz,
  ``The local B-polarization of the CMB: a very sensitive probe of cosmic
  defects,''
 Phys.\ Lett.\ B {\bf 695}, 26 (2011)
  [arXiv:1003.0299 [astro-ph.CO]].
  %%CITATION = ARXIV:1003.0299;%%

\bibitem{CSsimuls}
A.~Albrecht and N.~Turok,
  ``Evolution Of Cosmic Strings,''
  Phys.\ Rev.\ Lett.\  {\bf 54}, 1868 (1985);\\
  %%CITATION = PRLTA,54,1868;%%
D.~P.~Bennett and F.~R.~Bouchet,
  ``Evidence For A Scaling Solution In Cosmic String Evolution,''
  Phys.\ Rev.\ Lett.\  {\bf 60}, 257 (1988);\\
  %%CITATION = PRLTA,60,257;%%
B.~Allen and E.~P.~S.~Shellard,
  ``Cosmic String Evolution: A Numerical Simulation,''
  Phys.\ Rev.\ Lett.\  {\bf 64}, 119 (1990);\\
  %%CITATION = PRLTA,64,119;%%
C.~Ringeval, M.~Sakellariadou and F.~Bouchet,
  ``Cosmological evolution of cosmic string loops,''
  JCAP {\bf 0702}, 023 (2007)
  [arXiv:astro-ph/0511646];\\
  %%CITATION = JCAPA,0702,023;%%  
V.~Vanchurin, K.~D.~Olum and A.~Vilenkin,
  ``Scaling of cosmic string loops,''
  Phys.\ Rev.\  D {\bf 74}, 063527 (2006)
  [arXiv:gr-qc/0511159];\\
  %%CITATION = PHRVA,D74,063527;%%
J.~J.~Blanco-Pillado, K.~D.~Olum and B.~Shlaer,
  ``Large parallel cosmic string simulations: New results on loop production,''
  Phys.\ Rev.\ D {\bf 83}, 083514 (2011)
  [arXiv:1101.5173 [astro-ph.CO]].
  %%CITATION = ARXIV:1101.5173;%%

\bibitem{SPTPol}
J.~E.~Austermann, K.~A.~Aird, J.~A.~Beall, D.~Becker, A.~Bender, B.~A.~Benson, L.~E.~Bleem and J.~Britton {\it et al.},
  ``SPTpol: an instrument for CMB polarization measurements with the South Pole Telescope,''
  Proc.\ SPIE Int.\ Soc.\ Opt.\ Eng.\  {\bf 8452}, 84520E (2012)
  [arXiv:1210.4970 [astro-ph.IM]].
  %%CITATION = ARXIV:1210.4970;%%

\bibitem{ACTPol}
M.~D.~Niemack, P.~A.~R.~Ade, J.~Aguirre, F.~Barrientos, J.~A.~Beall, J.~R.~Bond, J.~Britton and H.~M.~Cho {\it et al.},
  ``ACTPol: A polarization-sensitive receiver for the Atacama Cosmology Telescope,''
  Proc.\ SPIE Int.\ Soc.\ Opt.\ Eng.\  {\bf 7741}, 77411S (2010)
  [arXiv:1006.5049 [astro-ph.IM]].
  %%CITATION = ARXIV:1006.5049;%%

\bibitem{flatsky}
M.~Zaldarriaga and U.~Seljak,
  ``An all sky analysis of polarization in the microwave background,''
  Phys.\ Rev.\ D {\bf 55}, 1830 (1997)
  [astro-ph/9609170];\\
  %%CITATION = ASTRO-PH/9609170;%%
W.~Hu,
  ``Weak lensing of the CMB: A harmonic approach,''
  Phys.\ Rev.\ D {\bf 62}, 043007 (2000)
  [astro-ph/0001303].
  %%CITATION = ASTRO-PH/0001303;%%

\bibitem{deficit}
A.~Vilenkin,
  ``Gravitational Field Of Vacuum Domain Walls And Strings,''
  Phys.\ Rev.\  D {\bf 23}, 852 (1981).
  %%CITATION = PHRVA,D23,852;%%

\bibitem{wakeZel}
R.~H.~Brandenberger, L.~Perivolaropoulos and A.~Stebbins,
  ``Cosmic Strings, Hot Dark Matter and the Large-Scale Structure of the Universe,"
  Int.\ J.\ Mod.\ Phys.\  A {\bf 5}, 1633 (1990);\\
  %%CITATION = IMPAE,A5,1633;%%
L.~Perivolaropoulos, R.~H.~Brandenberger and A.~Stebbins,
  ``Dissipationless Clustering Of Neutrinos In Cosmic String Induced Wakes,''
  Phys.\ Rev.\  D {\bf 41}, 1764 (1990).
  %%CITATION = PHRVA,D41,1764;%% 

\bibitem{Zel}
Y.~.B.~Zeldovich,
  ``Gravitational instability: An Approximate theory for large density perturbations,''
  Astron.\ Astrophys.\  {\bf 5}, 84 (1970).
  %%CITATION = AAEJA,5,84;%%

\bibitem{Bunn}
E.~F.~Bunn, M.~Zaldarriaga, M.~Tegmark and A.~d.~Oliveira-Costa,
  ``E/B decomposition of finite pixelized CMB maps,''
  Phys.\ Rev.\ D {\bf 67}, 023501 (2003)
  [astro-ph/0207338];\\
  %%CITATION = ASTRO-PH/0207338;%%
K.~M.~Smith and M.~Zaldarriaga,
  ``A general solution to the E-B mixing problem,''
  Phys.\ Rev.\ D {\bf 76}, 043001 (2007)
  [astro-ph/0610059].
  %%CITATION = ASTRO-PH/0610059;%%

\bibitem{Nick} N. Park, McGill Univ. PhD thesis, in preparation (2013).

\bibitem{Grant} G. Salton, McGill Univ. MSc thesis, 2013.

\bibitem{SPT}
 J.~E.~Ruhl {\it et al.}  [The SPT Collaboration],
  ``The South Pole Telescope,''
  Proc.\ SPIE Int.\ Soc.\ Opt.\ Eng.\  {\bf 5498}, 11 (2004)
  [arXiv:astro-ph/0411122].
  %%CITATION = PSISD,5498,11;%%
 
\bibitem{ACT}
A.~Kosowsky  [the ACT Collaboration],
  ``The Atacama Cosmology Telescope Project: A Progress Report,''
  New Astron.\ Rev.\  {\bf 50}, 969 (2006)
  [arXiv:astro-ph/0608549].
  %%CITATION = ASTRE,50,969;%% 

\bibitem{Gil}
M.~Kaplinghat, M.~Chu, Z.~Haiman, G.~Holder, L.~Knox and C.~Skordis,
  ``Probing the Reionization History of the Universe using the Cosmic Microwave
  Background Polarization,''
  Astrophys.\ J.\  {\bf 583}, 24 (2003)
  [arXiv:astro-ph/0207591].
  %%CITATION = ASJOA,583,24;%%

\bibitem{Canny}
J. Canny, 
  ``A computational approach to edge detection'',
  IEEE Trans. Pattern Analysis and Machine Intelligence {\bf 8}, 679 (1986).

\bibitem{Danos}
S.~Amsel, J.~Berger and R.~H.~Brandenberger,
  ``Detecting Cosmic Strings in the CMB with the Canny Algorithm,''
  JCAP {\bf 0804}, 015 (2008)
  [arXiv:0709.0982 [astro-ph]];\\
  %%CITATION = JCAPA,0804,015;%%
A.~Stewart and R.~Brandenberger,
  ``Edge Detection, Cosmic Strings and the South Pole Telescope,''
  arXiv:0809.0865 [astro-ph];\\
  %%CITATION = ARXIV:0809.0865;%%
R.~J.~Danos and R.~H.~Brandenberger,
  ``Canny Algorithm, Cosmic Strings and the Cosmic Microwave Background,''
  arXiv:0811.2004 [astro-ph];\\
  %%CITATION = ARXIV:0811.2004;%%
R.~J.~Danos and R.~H.~Brandenberger,
  ``Searching for Signatures of Cosmic Superstrings in the CMB,''
  arXiv:0910.5722 [astro-ph.CO].
  %%CITATION = ARXIV:0910.5722;%% 


\end{thebibliography}
\end{document}